%% file: main.tex
\documentclass[sigconf]{acmart}

\usepackage{enumitem}

\AtBeginDocument{%
  \providecommand\BibTeX{{%
    \normalfont B\kern-0.5em{\scshape i\kern-0.25em b}\kern-0.8em\TeX}}}
    
\copyrightyear{2020}
\acmYear{2020}
\setcopyright{acmlicensed}
\acmConference[ICSE-SEIP '20]{Software Engineering in Practice}{May 23--29, 2020}{Seoul, Republic of Korea}
\acmBooktitle{Software Engineering in Practice (ICSE-SEIP '20), May 23--29, 2020, Seoul, Republic of Korea}
\acmPrice{15.00}
\acmDOI{10.1145/3377813.3381353}
\acmISBN{978-1-4503-7123-0/20/05}

\begin{document}

%
% The "title" command has an optional parameter, allowing the author to define a "short title" to be used in page headers.
\title{DeCaf: Diagnosing and Triaging Performance Issues in Large-Scale Cloud Services}

\author{Chetan Bansal}
\email{chetanb@microsoft.com}
\affiliation{%
  \institution{Microsoft Research}
  \streetaddress{}
  \city{Redmond}
  \state{WA, USA}
}

\author{Sundararajan Renganathan*}
\email{rsundararajan20@gmail.com}
\affiliation{%
  \institution{Stanford University}
  \streetaddress{}
  \city{Stanford}
  \state{CA, USA}
}

\author{Ashima Asudani, Olivier Midy}
\email{ashimaa,olmidy@microsoft.com}
\affiliation{%
  \institution{Microsoft}
  \streetaddress{}
  \city{Redmond}
  \state{WA, USA}
}

\author{Mathru Janakiraman*}
\email{mathruj@gmail.com}
\affiliation{%
  \institution{Amazon}
  \streetaddress{}
  \city{Seattle}
  \state{WA, USA}
}

\thanks{*Work done while at Microsoft}

%\thispagestyle{plain}
%\pagestyle{plain}

\begin{abstract}
Large scale cloud services use Key Performance Indicators (KPIs) for tracking and monitoring performance. They usually have Service Level Objectives (SLOs) baked into the customer agreements which are tied to these KPIs. Dependency failures, code bugs, infrastructure failures, and other problems can cause performance regressions. It is critical to minimize the time and manual effort in diagnosing and triaging such issues to reduce customer impact. Large volume of logs and mixed type of attributes (categorical, continuous) in the logs makes diagnosis of regressions non-trivial.

In this paper, we present the design, implementation and experience from building and deploying DeCaf, a system for automated diagnosis and triaging of KPI issues using service logs. It uses machine learning along with pattern mining to help service owners automatically root cause and triage performance issues. We present the learnings and results from case studies on two large scale cloud services in Microsoft where DeCaf successfully diagnosed 10 known and 31 unknown issues. DeCaf also automatically triages the identified issues by leveraging historical data. Our key insights are that for any such diagnosis tool to be effective in practice, it should a) scale to large volumes of service logs and attributes, b) support different types of KPIs and ranking functions, c) be integrated into the DevOps processes.
\end{abstract}


%
% By default, the full list of authors will be used in the page headers. Often, this list is too long, and will overlap
% other information printed in the page headers. This command allows the author to define a more concise list
% of authors' names for this purpose.
\renewcommand{\shortauthors}{Chetan Bansal, Sundararajan Renganathan, Ashima Asudani, Olivier Midy, and Mathru Janakiraman}

%
% This command processes the author and affiliation and title information and builds
% the first part of the formatted document.
\maketitle

\input{introduction.tex}
\input{challenges.tex}
\input{related.tex}
\input{overview.tex}
\input{approach.tex}
\input{implementation.tex}
\input{case_studies.tex}
\input{experiments.tex}
\input{discussion.tex}
\input{conclusion.tex}
\input{acknowledgements.tex}
%
% The next two lines define the bibliography style to be used, and the bibliography file.
\bibliographystyle{ACM-Reference-Format}
\bibliography{references}
\end{document}

%% file: introduction.tex
\section{Introduction}

The move from boxed software to cloud services has changed how these products are built and deployed. It has simplified critical aspects of software development like shipping updates and compatibility with client hardware. This has also introduced a new role of DevOps where the service owners are responsible and accountable for meeting Service Level Objectives (SLOs) on Key Performance Indicators (KPIs). Large scale cloud services companies like Amazon, Facebook, Google and Microsoft have 100s of cloud services powering consumer and enterprise apps and websites. These cloud services use KPIs like latency, failure rate, availability, uptime, etc. to continuously monitor service health and user satisfaction. For a lot of commercial services, meeting SLOs with respect to these KPIs is often baked into the customer contracts and tied to service revenue. For instance, Amazon AWS Compute gives 10\% service credit if uptime is less than 99.99\% and 30\% service credit if uptime is less than 99.0\% \cite{Amazon}. There can also be indirect impact of regressions in performance, for instance, a 400 ms increase in latency causes about 0.5\% drop in Google search volume \cite{SpeedMatters}.

Root causing and diagnosing performance issues in distributed systems is a well studied problem in the Systems and Software Engineering communities. Existing work on log based performance diagnosis for services mainly relies on either anomaly detection \cite{khanduja2015near, gabel2012latent} or association rule mining based methods \cite{brauckhoff2009anomaly, yairi2001fault}. However, DeCaf is not comparable to these methods because of several reasons. Anomaly detection methods cannot scale to high dimensional and high cardinality data. For instance, \cite{khanduja2015near, gabel2012latent} have used anomaly detection on \texttildelow{}100 numerical counters, while we have applied DeCaf on data with categorical attributes with up to \texttildelow{}1M cardinality. Similarly, association rule mining based methods are not applicable to data with continuous attributes and KPIs, for instance, latency. Also, high dimensional and cardinality data will lead to a combinatorial explosion. Further, anomaly detection based methods also fail to detect pre-existing performance issues. We designed and built DeCaf with these limitations in mind. It is an end-to-end system for diagnosing and triaging performance issues in large scale services. We deployed and integrated it into the DevOps processes for 2 large scale cloud services in Microsoft, where it was able to successfully diagnose 31 unknown issues.

\begin{figure*}
\includegraphics[width=\textwidth]{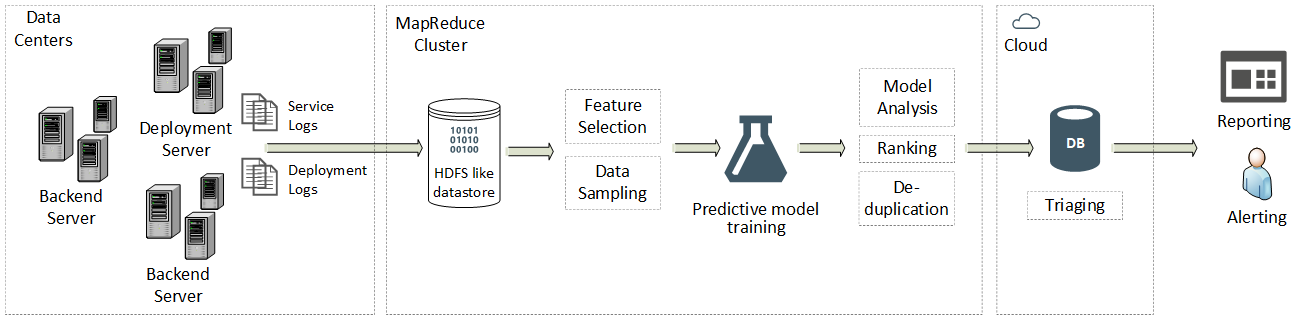}
\caption{Diagnosis and triaging workflow in DeCaf}
\label{figure:1}
\end{figure*}

\textbf{Contributions}: In this work, we designed and implemented DeCaf, a generic system for automated diagnosis and triaging of KPI issues. Figure \ref{figure:1} shows the overall workflow of DeCaf. Using existing service logs, it is able to diagnose known and unknown issues in 2 large scale services resulting in significant time savings and minimizing customer impact. It is also able to handle large volumes and large cardinality of logs as well as different types of KPIs. Lastly, it builds a knowledge base of results over time and can automatically triage newly detected issues. We share the results and learnings from deploying and evaluating DeCaf on 2 cloud services in Microsoft. We make the following contributions in this paper:

\begin{enumerate}[leftmargin=3mm,itemsep=0mm]
    \item We propose DeCaf, an end-to-end system for automatic diagnosis and triaging of performance issues in large scale cloud services from service logs.
    \item It introduces a novel approach which combines machine learning and pattern mining for diagnosing and triaging different types of KPI issues from large volume and high cardinality logs.
    \item We have integrated DeCaf into the DevOps processes of Microsoft starting from data collection to reporting and alerting DevOps engineers.
    \item We have deployed DeCaf on 2 large cloud scale services in Microsoft. The results confirm the usefulness of the system both in terms of diagnosing 41 known and unknown performance issues and, also, significantly reducing the manual effort in diagnosing performance issues.
\end{enumerate}
The rest of the paper is organized into following sections: In Section 2, we discuss the challenges in diagnosing and triaging of performance issues in large scale services. In Section 3, we discuss the related work. We provide an overview of the DeCaf system in Section 4. In Section 5, we describe our approach in detail. In Section 6, we describe the implementation details. In Section 7, we present the results from case studies on 2 large scale cloud services in Microsoft. In Section 8, we do an experimental evaluation of DeCaf for accuracy and runtime performance. We discuss the applicability along with future work in Section 9. Finally, we conclude with a summary in Section 10.

%% file: challenges.tex
\section{Challenges}

In Microsoft, we operate O(100) external cloud services powering several collaboration and cloud compute services running on O(100K) servers with O(10K) developers checking in code every week. This results in a huge amount of code, dependency and infrastructure churn which can lead to various kind of performance regressions. Diagnosing such issues using existing techniques is not only time consuming but also requires custom dashboards and manual investigations. Moreover, there could be important issues which are left undiscovered, as we show in our case studies. We studied several large scale cloud services in Microsoft and made the following observations regarding the diagnostic practices and challenges:

\begin{enumerate}[leftmargin=3mm,itemsep=0mm]
    \item \textbf{Logging}: These services log key request attributes and metrics. The attributes are of mixed types (categorical, continuous) and contain both structured and unstructured information. Attributes which are useful for monitoring and large-scale diagnosis (such as backend server, component latencies, etc.) are logged in a structured manner. Information like exception stack traces used for diagnosing request level problems are serialized and logged in an unstructured format. These services produce massive volumes of logs, up to O(1TB) per hour. So, the logs are aggregated periodically and stored in a central HDFS like massive data store with a Hadoop like map reduce system for data analytics.
    \item \textbf{Performance regressions}: Performance degradation can be due to various reasons, such as code bugs, infrastructure failure or overload, design gaps and dependency failures. These can also be local or global depending on various factors such as the root cause, deployment scope, etc. For instance, a hardware failure in a data center can increase the load on the other servers, degrading the request latency for that data center. Similarly, a thread contention bug can impact the performance of the entire service.
    \item \textbf{Diagnosis}: DevOps engineers usually do performance diagnosis for two purposes:
    \begin{enumerate}
        \item \textbf{Reactive}: To resolve an issue discovered through active monitoring or customer complaints.
        \item \textbf{Proactive}: To proactively improve the service performance and address performance gaps / bugs in the current codebase.
    \end{enumerate}
    For both proactive and reactive diagnosis, the service owners use custom KPI dashboards and the logs. Based on the issue, DevOps engineers might use the dashboard to manually narrow down the scope of the regression. Subsequently, they run custom queries to extract the request logs and, also, determine the impact of the issue. Sometimes, the diagnosis can also require additional logging or gathering performance counters from other sources.
\end{enumerate}

\textbf{Challenges}: We made the following key observations about the challenges faced in diagnosing performance issues in large scale cloud services:

\begin{enumerate}[leftmargin=3mm,itemsep=0mm]
    \item \textbf{Triaging} - Triaging is the process of prioritizing and determining which issue should be investigated and subsequently fixed. This is critical since it can take anywhere from hours to days to root cause and fix an issue. DevOps engineers consider various factors while triaging issues:
    \begin{enumerate}
        \item What is the impact of the issue? How many customers or requests are being impacted?
        \item Is the issue localized or global?
        \item Is it a known issue? If yes, has the impact increased?
        \item Has this issue occurred in the past?
    \end{enumerate} 
    Triaging performance issues is non-trivial because engineers have to estimate the scope and impact of issues by taking into account historical data.
    \item \textbf{Volume of logs} - A large scale service can generate anywhere from gigabytes to 100s of terabytes of logs every day. Manual querying and processing of logs for root causing issues is not scalable. Also, it is very expensive in terms of time and space to load such large amount of data in tools like Excel or in an SQL database.
    \item \textbf{High cardinality attributes} - Log attributes from large scale services usually have high cardinality, making it infeasible to manually explore various combinations for root causing issues. For instance, in the Orion service, which is one of the case studies for DeCaf, there are more than O(1M) organizations. The KPIs can regress for a subset of these organizations which makes it important to include such high cardinality data in diagnosing regressions. In Table \ref{table:attributes}, we list the type of cardinality of the some of the attributes from the Orion logs.
    \item \textbf{Mixed attributes and KPI labels} - Logs can contain both categorical and continuous features. For instance, in the Orion service logs, we have both categorical features like backend, front end machines, and continuous features like sub-component latencies.
    
    Similarly, the log attributes for KPIs can also be categorical and continuous. While latency is a continuous metric, request status is a binary categorical variable (success or fail).
    \item \textbf{Interpretability} - It is easier for DevOps engineers to investigate issues if the scope, performance impact and historical data is provided. One of the key insights which enabled us to successfully deploy DeCaf to multiple services within Microsoft is to make the results interpretable both quantitatively and qualitatively.
\end{enumerate}

\begin{table}
\begin{center}
 \begin{tabular}{|l|l|l|}
 \hline
 \textbf{Attribute} & \textbf{Type} & \textbf{Cardinality} \\
 \hline
 Organization & Categorical & \textasciitilde1M \\
 \hline
 Server & Categorical & \textasciitilde10K \\
 \hline 
 ApiPath & Categorical & \textasciitilde1K \\
 \hline 
 AppCategory & Categorical & \textasciitilde10 \\
 \hline 
 AppId & Categorical & \textasciitilde100 \\
 \hline
 CapacityUnit & Categorical & \textasciitilde100 \\
 \hline 
 UserAgent & Categorical & \textasciitilde100 \\
 \hline 
 DataCenterTarget & Categorical & \textasciitilde10 \\
 \hline 
 DataCenterOrigin & Categorical & \textasciitilde10 \\
 \hline 
 Forest & Categorical & \textasciitilde10 \\
 \hline 
 UserCategory & Categorical & \textasciitilde10 \\
 \hline 
 BuildVersion & Categorical & \textasciitilde10 \\
 \hline 
 UserCache L1 Latency & Continuous & - \\
 \hline 
 L2 Cache Latency & Continuous & - \\
 \hline 
 LoadBalancer Latency & Continuous & - \\
 \hline
 Auth Latency & Continuous & - \\
 \hline 
\end{tabular}
\end{center}
\vspace{10pt}
\caption{Sample attributes from the Orion logs}
\vspace{-20pt}
\label{table:attributes}
\end{table}

%% file: related.tex
\section{Related work}

Leveraging machine learning techniques to perform diagnostics on service logs has been the focus of much research over the past couple of decades \cite{bodik2010fingerprinting, yim2016evaluation, ChenICAC}. Bodik et. al \cite{bodik2010fingerprinting} rely on anomaly signatures of known issues along with regression models for diagnosing failures in data centers. Chen et. al \cite{ChenICAC}, use classification trees to root cause failure rates in a large internet website like eBay. Cohen et. al \cite{cohen2004correlating} use Tree-Augmented Bayesian Networks to identify combinations of system level metrics which are correlated with non-compliance of SLOs. Nair et. al \cite{nair2015learning} using hierarchical detectors with time series anomaly detection to diagnose issues. A combination of clustering and anomaly detection techniques for root causing has also been proposed \cite{duan2009fa, farshchi2015experience}. However, these methods are not feasible for high cardinality data. We distinguish our work by proposing a simple end-to-end system which can a) handle heterogeneous and high cardinality O(1M) data, b) diagnose different types of KPIs, c) can automatically rank and triage the discovered issues, d) detect previously unknown issues.

A lot of prior work has also focused on analyzing raw service logs to extract meaningful events and diagnose abnormal system behavior. Xu et. al \cite{xu2009detecting} jointly analyze source code and console logs to extract features and perform anomaly detecting on these feature collections. Deeplog \cite{du2017deeplog} models the sequence of events producing log files using LSTMs and constructs workflows to aid in root causing when it is inferred that the log patterns have deviated from the trained model. LogCluster \cite{vaarandi2015logcluster} groups together log messages to construct representative log sequences thereby assisting engineers in diagnosing failures. Distalyzer \cite{nagaraj2012structured} consumes two sets of logs, one with good performance and one with bad performance to extract systems behaviors that diverge the most across the two sets of logs and are correlated with bad system performance. AUDIT \cite{luo2018troubleshooting} takes a slightly orthogonal path by setting up lightweight triggers to identify the first instance of a problem and then uses blame-proportional logging to when the problem reoccurs. Zawawy et. al \cite{zawawy2010log} propose a log reduction framework which filters and interprets a subset of streaming log data in order to perform root cause analysis. In this work, we diagnose issues by using existing structured data logged by cloud services. It is a reasonable constraint to work under because most large-scale services log structured data to simplify monitoring, debugging and analytics.

The empirical software engineering community has done a lot of work on automated bug triaging and characterization. Tian et al. \cite{tian2012improved} and Alenezi et al. \cite{alenezi2013efficient} use textual features from bug reports for identification of duplicate bug reports. Xia et al. \cite{xia2017improving} uses topic modelling to assign the bug re-ports to the appropriate developer. Lamkanfi et al. \cite{lamkanfi2010predicting} extract textual features from the bug reports to predict the severity of the bug reports to assist in triaging. These techniques are complementary to DeCaf since they are applicable only after a detailed bug report has been filed in the bug tracker. To the best of our knowledge, DeCaf is the first end-to-end system which uses historical results for triaging the issues.

%% file: overview.tex
\section{Overview}
One of the major guiding principles while designing DeCaf was to build a generic system using simple interpretable techniques so that service owners can themselves own and deploy DeCaf. We have deployed DeCaf to two large scale services in Microsoft, Orion and Domino (name changed):
\begin{enumerate}[leftmargin=3mm,itemsep=0mm]
    \item Orion - Orion is a stateless routing service that proxies requests to the correct backend / mailbox server (the one hosting the user's active mailbox database). It handles about O(100B) requests every day and logs about O(100) attributes per request, generating O(100) terabytes of logs daily. Request latency is a key metric in Orion which impacts the latency of all clients.
    \item Domino - Domino is a global scale internet measurement platform built for one of the major cloud providers. It is designed to perform client-to-cloud path measurement from users around the world to the first-party and third-party endpoints. Failure rate is one of the KPIs for Domino since it is tied to service availability. 
\end{enumerate}
\textbf{Goal}: With DeCaf, our goal was to build a system which can automatically help the DevOps engineers narrow down a performance regression to a subset of requests. So, for each identified issue, it outputs a root-cause which consist of a set of predicates along with impact metrics and a triage category. These predicates help narrow down a regression to a subset of log rows and columns which can then be further used for mitigation of the issue. A predicate is a boolean valued function defined as:
\\
\begin{equation}
P(X) \rightarrow{} \{true, false\}
\end{equation}
\\
Predicates can be defined for both continuous and categorical log attributes, for example:
\\
\begin{equation}
AuthLatency > 50ms \rightarrow{} \{true, false\} 
\end{equation}
\begin{equation}
Region = NorthAmerica \rightarrow{} \{true, false\}
\end{equation}
\\
A DeCaf result consists of the following data:
\begin{itemize}[leftmargin=3mm,itemsep=0mm]
    \item \textbf{Correlated predicate}: Predicate correlated with performance regression.
    \item \textbf{Scope predicates}: Predicates which define the scope of the regression.
    \item \textbf{Count} of impacted requests in sampled data.
    \item \textbf{Performance impact}: Impact on the KPI.
    \item \textbf{Triage result (r)}: r $\in$ \{new, existing, regressed, improved, resolved\}
\end{itemize}
\textbf{Example}: To better understand the problem and the goals, here is an actual incident from the Orion service which was root caused using DeCaf: \\
In the Orion service, DeCaf discovered a latency regression in the AN150C01 server rack which was causing ~90,000\% higher latency than SLO for requests from offbox requests. On investigation, it was root caused to an auth component which was not being logged and was causing timeouts and errors. It was impacting the latency of \textasciitilde{}1 Billion requests worldwide daily. Here is the output of DeCaf from this incident:
\begin{itemize}[leftmargin=3mm,itemsep=0mm]
    \item \textbf{Correlated predicate}: Rack:AN150C01
    \item \textbf{Scope predicates}: \{RequestType:Offbox $\land$ LocDataCenter:AN $\land$ CrossDataCenter:true\}
    \item \textbf{Requests impacted}: \textasciitilde{}1 Billion
    \item \textbf{Latency impact}: 4419ms
    \item \textbf{Triage result}: new
\end{itemize}

%% file: approach.tex
\section{Our Approach}
In this section, we describe the details of DeCaf, which consists of 4 major steps: data preparation from the raw service logs, training of ML models, rule extraction from the learnt models and triaging of issues based on historical data. Out of these 4 steps, only the data preparation and model training steps require one time manual effort while deploying DeCaf for the first time.
\subsection{Data preparation}
Data sampling and feature selection is key to any machine learning or data-mining based system. Cloud services can generate 100s of terabytes of data every day and log hundreds of attributes. This makes any manual or automated analysis challenging both in terms of space and time. To solve this problem, we leverage data sampling and feature selection. 
\subsubsection{\textbf{Data pre-processing}} Often the service logs are distributed across multiple data streams. For instance, different components of a service can write logs to different data stores while logging a distinct correlation Id per request to help join the logs at a later stage. So, as a one time step, service owners write queries for aggregating / joining data from various streams/sources. Similarly, often the data is serialized while being logged, so, we also de-serialize the data into a structured schema. To handle missing values, we replace any missing value in categorical attributes by a placeholder ("<EMPTY>") while for continuous attributes we use the median value.
\subsubsection{\textbf{Feature selection}} Not all features present in the service logs are useful for root causing issues. For instance, features such as Request Id or Correlation Id which uniquely identifies individual rows are not useful in root causing widespread issues. These features can also have very high cardinality and can cause state explosion. To be able to select the right features, we follow a hybrid approach:
\begin{itemize}[leftmargin=3mm,itemsep=0mm]
    \item DeCaf automatically classifies features into continuous and categorical types and measures the cardinality of the categorical variables. It provides recommendations for pruning the high cardinality features.
    \item Service owners further prune the features based on their domain knowledge and the recommendations provided by DeCaf.
\end{itemize}
\subsubsection{\textbf{Stratification}} Stratification is the process of dividing the data into mutually exclusive, homogeneous and collectively exhaustive subsets or classes. While it is possible to have multiple classes in stratification, we only consider binary classes in this work since it supports diagnosis of all the KPIs we have seen in practice. 
We divide the service requests into two classes:
\begin{enumerate}[leftmargin=3mm,itemsep=0mm]
    \item Positive: Requests showing anomalous behavior or violating the performance SLO for the KPIs.
    \item Negative: Requests which meet the KPI SLOs.
\end{enumerate}
The stratification criteria is determined by the SLO and is an input to DeCaf. For instance, in Orion, based on the latency SLO of the service, the criteria is:
\begin{equation}
C = \begin{cases}
  positive, & RequestLatency > 5ms, \\
  negative, & RequestLatency \leq 5ms.
\end{cases}
\end{equation}
\subsubsection{\textbf{Sampling}} The Orion and Domino services generate more than 100 terabytes and 10 terabytes of logs every day, respectively. Using all this data will result in performance bottlenecks both in runtime and computational resources. There are several different statistical methods of sampling data: random sampling, systematic sampling, stratified sampling, cluster sampling, etc. In this work, we considered random and stratified sampling of data:
\begin{itemize}[leftmargin=3mm,itemsep=0mm]
    \item \textbf{Random sampling}: Data is randomly selected from the entire population.
    \item \textbf{Stratified sampling}: Data from the positive and negative subsets is randomly sampled separately within each of the strata. 
\end{itemize}
We selected the sampling method based on two empirical observations:
\begin{enumerate}[leftmargin=3mm,itemsep=0mm]
    \item \textbf{Class imbalance}: Most production services operate within SLO requirements most of the time. For instance, we observed only 0.1\% requests missing the SLO in Orion.
    \item \textbf{Target variable}: Percentage metrics like failure rate are computed on the entire population, so, we do random sampling of data for such metrics. For absolute metrics like request latency, we use stratified sampling.
\end{enumerate}

Also, sampling the data improves efficiency in terms of runtime and compute resources.

\iffalse
\begin{figure*}[h]
\includegraphics[width=\textwidth]{figures/trees.png}
\caption{Example of a regression tree (left) and classification tree (right)}
\label{figure:2}
\end{figure*}
\fi

\subsection{ML model training}
Random forest models \cite{Breiman2001} have been effectively utilized for various tasks such as image processing \cite{rogez2008randomized}, churn prediction \cite{xie2009customer}, intrusion detection \cite{farnaaz2016random}. In DeCaf, we use Random Forest models for learning predicates which correlate with performance issues. It is an ensemble machine learning method for classification and regression that operates by constructing a multitude of decision trees \cite{quinlan1986induction}. For training the models, we use the KPI (latency, failure) as the target labels and the rest of the log attributes as the features or independent variables. Also, unlike conventional machine learning, we do not use the trained models for prediction since the goal is not to predict the KPI for future service requests. DeCaf analyzes the trained models to extract predicates which can help localize performance regressions.

\textbf{Decision trees}: A decision tree consists of a set of split and leaf nodes where each node is defined by a predicate. Essentially, decision trees are a hierarchy of nodes in the form of a tree. %Figure 2 shows a sample regression tree and classification tree. 
Decision trees can be used for both categorical and continuous target variables. If the target variable is categorical, classification trees are used; if it is continuous, regression trees are used. Both classification and regression trees follow similar process for training. At each training step, the best predicate is selected for partitioning the data and this process is repeated. The splitting criteria varies between classification and regression trees.

\textbf{Classification trees}: Classification trees are used to predict categorical target variables (for instance, weather-outlook: rain or sunny). At the training time, classification trees maximize the information gain at each split by reducing the entropy of the partitioned data after the split. Information gain when the tree is split on attribute $A_i$ is defined as:

\begin{equation}
IG(S,A_j) = H(S) - \sum_{A_i \in A} \frac{|A_i|}{S}H(A_i),
\end{equation}

where H(S), the entropy of set S, is defined as:

\begin{equation}
H(S) = - \sum_{c \in C} p_clog_2p_c,
\end{equation}

where $p_c$ is the probability of S for the class c.

\textbf{Regression trees}: Regression trees are used to predict continuous variables, for instance, temperature or age. Unlike classification trees, instead of maximizing information gain, regression trees minimize the mean squared error (MSE) at each split:

\begin{equation}
MSE = \frac{1}{n}\sum_{i=1}^{n}(y_i - \hat{y_i})^{2},
\end{equation}

where $y_i$ is the actual value of the target variable and $\hat{y_i}$ is the predicted value.

We use random forest ML models in DeCaf for several reasons:
\begin{enumerate}[leftmargin=3mm,itemsep=0mm]
    \item It is one of the few classes of models that can support not only continuous and categorical features but also continuous and categorical target variables.
    \item It is one of the most interpretable machine learning models \cite{huysmans2011empirical}.
    \item It is highly scalable both in terms of feature cardinality and data volume. Also, it is easily parallelizable on a MapReduce systems like Hadoop and Spark \cite{chen2017parallel, han2013scalable, li2012scalable}.
    \item As we evaluate in Section 6, it is less prone to overfitting and perform better than the decision tree baseline. The feature sampling ensures that strong predictive attributes in the data do not dominate all the trees.
\end{enumerate}

Random forest models, like other machine learning models, have several hyper-parameters which can be tuned to improve the prediction accuracy and runtime performance. Much work has been done to analyze the impact of the hyper-parameters on prediction accuracy \cite{bergstra2012random, bergstra2011algorithms}. However, in this work, instead of using the model for prediction, we use the trained models for extracting rules for diagnosis of KPI regressions. Based on empirical experiments, we found the following hyper-parameters to be useful:
\begin{enumerate}[leftmargin=3mm,itemsep=0mm]
    \item Min rows in leaves: Specifies the minimum number of training data in a leaf to avoid overfitting. If a leaf node contains less than this threshold, it will not continue to split the training data. The tree will stop growing on that leaf. This helps reduce the noise by eliminating rules which impact very small number of requests. Based on the discussion with the Orion and Domino teams, this was set to 1\% of the log size.
    \item Feature sample ratio: Specifies the sampling ratio used for sampling features when learning each tree. Setting the sampling ratio to 1 will cause all trees in the forest to be identical. Based, on empirical experiments we have set the sampling ratio to 0.6. 
    \item Number of trees: Number of trees to train. This increases the number of unique rules learnt for diagnosis while also increasing the training time. In DeCaf, based on runtime constraints, we train 50 trees.
\end{enumerate}
While onboarding a new service to DeCaf, a one time effort will be needed to tune the hyper-parameters based on the data schema, size and the compute resources. We will work on automating this manual step in future work.
	
\textbf{Model output}: After training the random forest model, we get a set of regression or classification trees. We dump the binary model into a text-based readable format. Each tree contains a hierarchical partitioned list of predicates starting from the root node to the leaf nodes. For each of the nodes, we also compute the following metrics:
\begin{enumerate}[leftmargin=3mm,itemsep=0mm]
    \item Row count: Number of training samples in a node.
    \item Anomaly probability (Classification trees): Probability of a training sample belonging to the positive (anomalous) class. 
    \item Predicted value (Regression trees): Average value of the target variable for the samples in a node.
\end{enumerate}
As we discuss in the next step, these metrics are used for ranking and triaging the results.

\iffalse
\begin{algorithm}
\begin{flushleft}
\textbf{Inputs:}\\
$F$: set of trained decision trees in random forest\\
$k$: delta extraction threshold\\
\textbf{Outputs:}\\
$P_i$: path predicate of potential root cause $R_i$\\
$m_i$: main predicate of root cause $R_i$\\
$I_i$: impact of $R_i$ on performance indicator\\
$N_i$: population size for $R_i$\\
$R$: set of rules extracted \\
$\mu_i$: average performance of affected population\\
% \ForEach {$t \in F $} {
\begin{algorithmic}
\Procedure{$extract\_rules$}{$t, cur\_node$, P}
%    \State $cur\_node \leftarrow t.root$
  \State $ left \leftarrow cur\_node.left $
  \State $ right \leftarrow cur\_node.right $
  \State $P \leftarrow P + cur\_node$
  \State $is\_left \leftarrow left - right > 0 $
  \State \textbf{if} ($\abs{is\_left} \geq k$)
  \State $regr\_true, regr\_false \leftarrow \text{regr}(left, right) $
%   \State $P, x = extract\_rules(t, y)$
   \State \text{return} ($P , is\_left, regr\_true), extract\_rules(t, regr\_false, P)$
\EndProcedure
\end{algorithmic}
\caption{DeCaf Rule Extraction Algorithm}
\label{alg:te}
\end{flushleft}
\end{algorithm}
\fi

\subsection{Rule extraction}
Once the random forest model is trained, we implement a novel algorithm to analyze the model output to produce a ranked list of rules as discussed in Section 4: 

\textbf{Step 1}: The random forest model generates a set of decision or regression trees. We parse the text output of the random forest into an in-memory set of decision tree objects and then recursively traverse each tree starting from the root node. At each node, we compute aggregate scores of the left and the right sub-trees. DeCaf exposes an interface so that the scoring function can be defined based on the requirements of the service owners using a lambda function. 

The scoring function takes in as parameters the metrics generated by the random forest model. The scoring function is defined by the service owners based on the SLO. For instance, in case of Orion, the SLO included not just the latency impact but also the number of requests impacted. Here are the scoring functions used in the Orion and Domino deployments:

\begin{itemize}[leftmargin=3mm,itemsep=0mm]
    \item Orion: Score(row count, predicted value) = row count x predicted value
    \item Domino: Score(row count, failure probability) = failure probability
\end{itemize}

\textbf{Step 2}: For each node predicate in the tree, we then compute a score for performance impact. It is the difference between the score of the left child node and the right child node, i.e., when the predicate is true vs when it is false:
\begin{equation}
Correlation Score = Score(LeftChild) - Score(RightChild)
\end{equation}
If the correlation score is positive, that means the predicate (P:X $\rightarrow{}$ true) is positively correlated with a performance degradation, otherwise, it is negatively correlated.

\textbf{Step 3}: Using the above algorithm, we extract a set of rules from the random forest model; where a rule consists of: (a)
Correlated predicate: predicate of the current node, (b)
Scope predicates: logical conjunction of the predicates from the parent nodes, (c) Correlation score and (d) Request count: No. of requests impacted.
	
\textbf{Step 4}: Next, we de-duplicate the rule set by only keeping the rules with the maximum impact score for each correlated predicate. Also, based on the feedback from the service owners, we remove the rules with negatively correlated predicates as they were not deemed useful by the service teams in root causing. For instance: 
ClientRegion:NorthAmerica$\rightarrow{}$False
only tells us that the client can be anywhere but in North America. While there might be cases where negative correlations are useful, we have not observed it in practice.

\subsection{Triaging} 
The results from the DeCaf algorithm are uploaded to an SQL database which allows us to build a historical knowledge base of all the detected issues. We leverage this history to automatically triage rules generated by DeCaf into the following categories based on the correlated predicate and the correlation score:
\begin{enumerate}[leftmargin=3mm,itemsep=0mm]
    \item \textbf{New}: A new predicate is extracted which has not appeared in the last 14 days.
    \item \textbf{Regressed}: Score of the predicate is more than 1 standard deviation above the mean score computed over the previous 14 days.
    \item \textbf{Known}: Score is within 1 standard deviation of the mean score computed over the previous 14 days.
    \item \textbf{Improved}: Score is at least 1 standard deviation below than the mean score computed over the previous 14 days.
    \item \textbf{Resolved}: Correlated predicates which were extracted in the previous day are not extracted.
\end{enumerate}
The regression and history thresholds were determined based on inputs from the service owners and empirical validation on 1 month of Orion data. To avoid the cold start problem, the auto-classification is enabled only once we have a 14 day history.

%% file: implementation.tex
\section{Implementation}
\begin{figure*}[h]
\includegraphics[width=\textwidth]{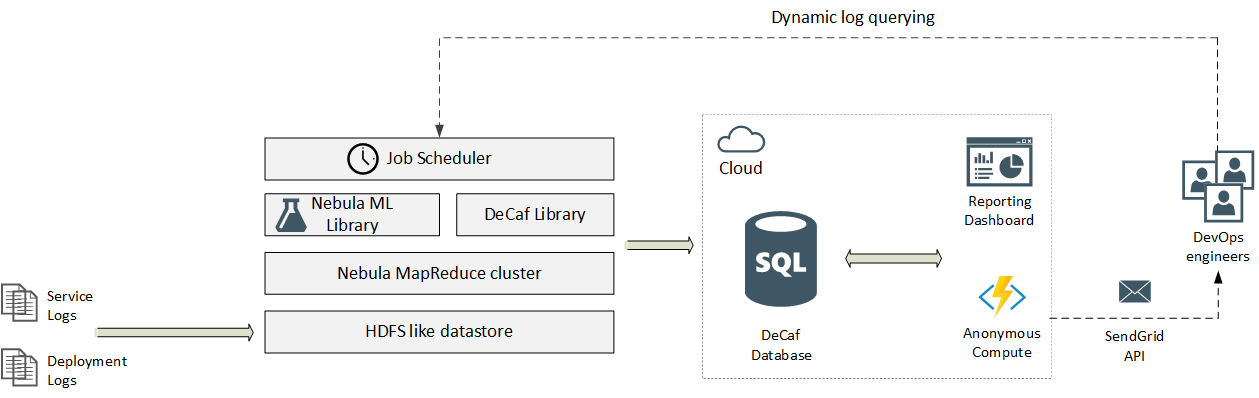}
\caption{DeCaf implementation}
\label{implementation}
\end{figure*}
We have implemented DeCaf using Microsoft Azure Cloud and a Hadoop like MapReduce cluster in Microsoft for large scale data analytics, called Cosmos. The overall architecture of DeCaf is shown in Figure \ref{implementation}. Cosmos supports an SQL like query language for running MapReduce jobs called Scope. We wrote modules using this query language for the data sampling and model training. Also, we implemented the rule extraction and triaging module using C\# (.NET Framework v4.5) and SQL. DeCaf is operationalized in Microsoft both for retrospective analysis, i.e., diagnosing regressions in the past and, also, for near real time analysis where service owners root cause ongoing incidents which were caught using alerts or customer complaints.\looseness=-1

\textbf{Data ingestion}: Large scale cloud services like Orion and Domino generate \textasciitilde100 TB of logs daily which are initially stored in the local storage of each server. A data loader process runs at scheduled intervals and scrubs the users' Personally Identifiable Information (PII) from these logs and uploads these raw logs to an HDFS like data store used by Cosmos. These logs are then processed by custom MapReduce jobs by the respective service owners for various purposes like analytics, monitoring and debugging. We use these processed logs in DeCaf.

We run the data ingestion job using a job scheduler for Cosmos called Avocado. Avocado also allows us specify dependency between various jobs and, also, enables us to visualize the results of the map-reduce tasks using a declarative JavaScript based frontend interface. The cadence depends on the service requirements, it can be in near real time or at fixed time intervals, like hourly. Sometimes logs from multiple sources might also be aggregated for diagnostics. For instance, in Orion, we used not only the Orion request logs but also infrastructure logs to be able to root cause performance issues related to infrastructure failures. 

\textbf{Model training and analysis}: For the Random Forest model training, we use an existing ML library for Cosmos available internally at Microsoft. It implements a distributed version of the CART algorithm for training Random Forest models. Similar distributed implementations are available for MapReduce systems like Hadoop and Spark \cite{chen2017parallel, han2013scalable, li2012scalable}.

We implemented a custom library in C\# .NET framework 4.5 for analyzing the random forest model output and generating the ranked list of rules. The library uploads the results to a Azure SQL cloud database. The Avocado job for model training and analysis is triggered once the data is ingested. To avoid having race conditions between the data and the model jobs, we have sequential dependency between the two jobs.
\begin{figure}[h!]
\includegraphics[width=\linewidth]{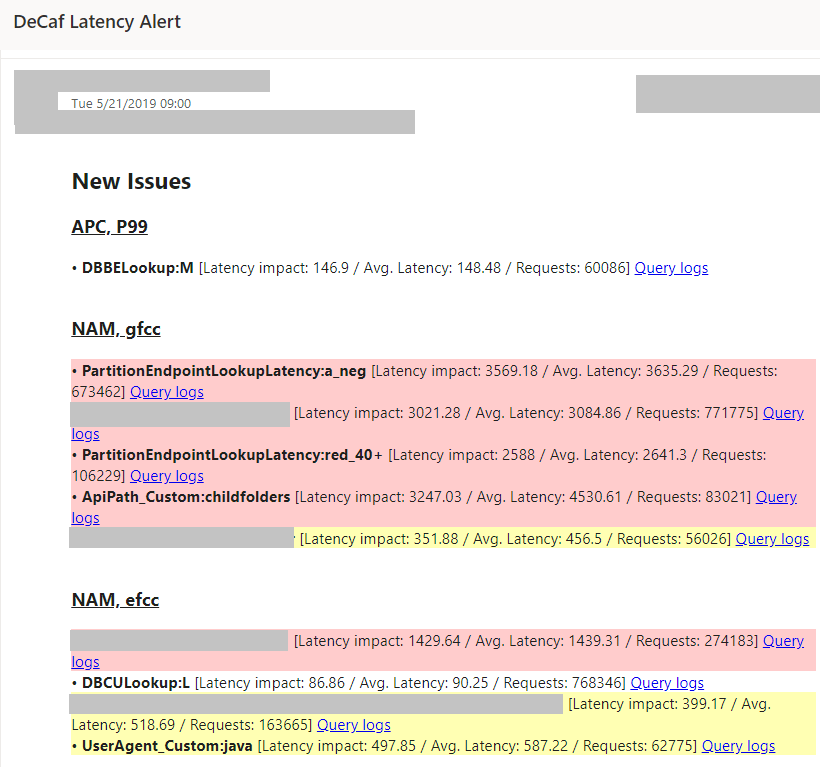}
\caption{DeCaf report for Orion}
\label{report}
\end{figure}

\textbf{DevOps integration}: To integrate DeCaf into the DevOps processes for the Orion and Domino services, we built a web dashboard, shown in Figure \ref{report},  which can be used to see the latest and historical results of diagnosis and triaging. However, based on developer feedback, we realized that a push-based notification mechanism is better suited than a pull-based one. Therefore, we also created a notification service which sends email notifications with the results using the SendGrid SMTP service.

\textbf{Dynamic query generation}: Before DeCaf was deployed, DevOps engineers had to manually write MapReduce scripts for mining the service logs for investigating service incidents. To reduce the manual effort, We implemented dynamic Cosmos MapReduce query generation for the rules generated by DeCaf. Each rule has a "Query Logs" link, as shown in Figure \ref{report}, which automatically runs a Cosmos job for mining the logs satisfying that rule. Based on the feedback from the Orion and Domino teams, it has significantly reduced investigation time and effort for incidents.

%% file: case_studies.tex
\section{Case Studies}
In this section, we present the results and learnings from deploying DeCaf on Orion and Domino. In Orion, we deployed DeCaf in production for 4 months, where the DevOps engineers used it to diagnose and triage issues impacting Orion latency in production. For Domino, we did a 10-day pilot where DeCaf was used to diagnose high failure rates.

\begin{table*}[t]
 \begin{tabular}{|l|p{3.5cm}|p{5cm}|l|p{5cm}|}
 \hline
 \textbf{No.} & \textbf{Correlated predicate} & \textbf{Scope predicates} & \textbf{Triage category} & \textbf{Issue description} \\
 \hline
 1 & CapacityUnit: CX20A1 & Region:NorthAmerica $\land$ AppName:AnonymizedApp $\land$ LocDataCenter:CX2 $\land$ CrossDatacenter:True & New & Gap in telemetry, latency of Auth token extraction was not being logged. \\
 \hline
 2 & RoutingLatency $>$ 568ms & Region:NorthAmerica & Regression & Shard based routing cross process calls to routing service hang in RPC layer \\ 
 \hline
 3 & UserLookup: L2 & Region:AsiaPacific $\land$ 
MailboxCacheLatency $>$ 40 $\land$
CapacityUnit:Anonymized & New & Regression in Asia Pacific region due to large no. of requests with cache miss \\
 \hline
 4 & RoutingLatency $>$ 16145ms & Forest: FC3 $\land$
CapacityUnit:CF5015A & Regression & Regression in Europe region due to lock contention bug in code \\
 \hline
 5 & UserMapping Cache Latency $>$ 132ms & UserLookup:L1 $\land$ Region:NorthAmerica & New & High L1 cache latency due to DB page misses and disk IO latency \\
 \hline
 6 & ServerLookup Cache: L2 & Forest: FA4 $\land$ Region: Europe & New & Large no. of L1 cache misses for backend server lookup \\
 \hline
 7 & AuthLatency Latency $>$ 47ms & AppName:AnonymizedApp $\land$ AddressResolution $<$ 5ms & New & High latency of requests due to timeout of the auth service \\
 \hline 
\end{tabular}
\vspace{10pt}
\caption{Some of the unknown issues diagnosed in Orion using DeCaf}
\vspace{-15pt}
\label{table:issues}
\end{table*}

\subsection{Orion}
Office365 is a large commercial collaboration and email service. It employs a horizontal scale-out architecture based on sharding by users across O(100K) servers across the globe that have both compute and storage capacity. The basic transaction processing model is to take compute as close to the data as possible, often to the very same server. Hence, we need a system to route requests efficiently to the specific server where the primary shard targeted by the request is presently activated.

Orion is the system that does this smart request routing as the request routing plane. Orion runs on an IIS web service and has multiple dependencies on internal as well as external sub-systems like shared caches, auth components, microservices, databases, etc. It is a massively distributed service that currently sustains a peak throughput of O(1M) requests/second, that target O(1B) shards, provisioned across O(100K) Office365 servers, spread over O(100) data-centers worldwide. Office365 has 1st party, 2nd party and 3rd party partners. Each of these partners have their own SLOs, and it becomes critical for the routing application layer to resolve the shard location correctly and route requests with SLO of latency less than RTT (round trip time) + 5ms at the 99th percentile. Before a user request lands on its shard\textquotesingle{}s location, it gets processed by multiple routing applications like load balancer, network layer and multiple hops in request routing.

The goal of using an automated root causing system for Orion was to not only detect regression but also find existing bugs and design flaws resulting in high latency. Given that Orion service generates O(100TB) of diagnostics data daily, receives O(100B) requests per day and logs O(100) request attributes resulting in a cumulative cardinality of O(1M), it becomes challenging to perform root cause analysis on latency regression through manual or traditional methods like big data processing, custom pivot dashboards, etc. 
Previously, root cause analysis was done manually by running queries on Cosmos as well as by analyzing Power BI \cite{PowerBI} reports over aggregated data in a cloud SQL database. Various limitations prevented Orion developers from diagnosing and investigating latency regressions at scale:
\begin{enumerate}[leftmargin=3mm,itemsep=0mm]
    \item Reporting limitation of maximum 1 GB data in their custom dashboards.
    \item Refresh time out if data is huge which required manual intervention.
    \item Given there are O(100) attributes, it was highly inefficient to click through various pivots manually for root causing.
\end{enumerate}
\textbf{Results}: We deployed DeCaf for root causing latency issues in Orion and discuss the results from a deployment done between March 2019 to September 2019. During this period, DeCaf was able to diagnose 9 known and 15 unknown issues. This significantly reduced the manual analysis overhead while also helping in finding new issues which would not have been detected otherwise. Table \ref{table:issues} lists the details of some of the unknown issues discovered in Orion impacting request latency. As per an analysis done by the Orion engineers, using DeCaf saved them on average 20 hours per investigation. Previously, most of this time was spent in finding the right predicates manually using dashboards and running MapReduce queries for validation and analysis.
\subsection{Domino}
Domino is a global scale internet measurement platform built for one of the major cloud providers. It is designed to perform client-to-cloud path measurements from users around the world to Microsoft\textquotesingle{}s first-party and third-party endpoints. The metrics it measures includes the availability of and the latency towards the above mentioned endpoints as seen by end-users spread across the globe. It is embedded in a variety of web-client and rich-client applications and performs measurements to the endpoints mentioned in the configuration file that it fetches from a pre-specified web location. It is deployed in O(10) web clients and rich clients and monitors O(100) end-points. The user-base performing these measurements is spread across O(10,000) ISPs (mobile and non-mobile) and O(1000) metro areas. The measurement platform generates O(10TB) of data consisting of around O(1B) requests, per day. Each request contains O(10) attributes with a cumulative cardinality of O(1M).

Failure rate is an important KPI being tracked as it is tied to service availability. It is defined as the number of failed Domino requests divided by the number of attempted requests in each time bucket. The scale and diversity of measurements being performed often resulted in the failure rate behaving in unexpected ways. One such issue was that certain large-scale client networks were facing higher failure rate during day times (in local time for the clients) as compared to night times. To zone in on the pivots which were resulting in increased failure rates, we applied DeCaf to the Domino data, as the high request volume coupled with the high cardinality of dimensions made manual analysis a problem of finding a needle in the haystack.

\textbf{Results}: In a 10-day pilot with Domino, DeCaf found 16 unknown issues and 1 known issue which were causing high failure rates. Out of the 16 unknown issues, 11 were related to specific tenants and 5 were related to certain Autonomous Systems (ASs). On further investigation, we found that these faulty tenants and ASs were facing more than 90\% failure rates. In addition to this, the fact that they performed more measurements during the day as compared to night times was resulting in a higher aggregate failure rate during day times, thereby explaining the initial observation. On following up with the respective service owners, we found that these high failure rates were caused by enterprise tenants blocking these endpoints in their firewalls. By filtering out these blocking tenants and ASs, as reported by DeCaf, we could eliminate the incidence of higher failure rate and reduce noise in the data.

%% file: experiments.tex
\section{Experimental evaluation}
\begin{table}
 \begin{tabular}{|c|c|c|} 
 \hline
 Algorithm & Precision & Valid Issues \\
 \hline
 Baseline & 0.66 & 2 \\
 \hline
 DeCaf & 0.72 & 8 \\  
 \hline
\end{tabular}
\vspace{10pt}
\caption{DeCaf evaluation}
\vspace{-15pt}
\label{comparison}
\end{table}
\begin{table}
 \begin{tabular}{|c|c|c|c|c|} 
 \hline
 Dataset & Pre-processing & Training & Diagnosing & Triaging \\
 \hline
 Small & 5 min & 12 min & 0.2 min & 0.8 min \\
 \hline
 Large & 138 min & 102 min & 0.3 min & 0.6 min \\
 \hline
\end{tabular}
\vspace{10pt}
\caption{Runtime comparison of DeCaf stages}
\vspace{-20pt}
\label{runtime-comparison}
\end{table}
For a quantitative evaluation, we compare the performance of DeCaf on 1 day of Domino service logs with the method proposed by Chen et al. \cite{ChenICAC}, referred to as the \textit{baseline}. We used 2 metrics for evaluation: precision and number of valid issues found. The results were manually evaluated by the service owners. Because of lack of ground truth of incidents and corresponding root causes in the test dataset, we did not evaluate recall. The metrics are defined as: \\
\textbf{Precision}:
        \begin{equation}
        Precision = \frac{|\{TP\}|}{|\{TP\}| + |\{FP\}|} 
        \end{equation}
\\
\textbf{Number of valid issues}: It is the number of valid issues discovered which were impacting performance. $|$\{TP\}$|$.

Here, \{TP\} is the set of correct results (true positives) and \{FP\} is the set of incorrect results (false positives).\\
As shown in Table \ref{comparison}, the number of valid issues discovered by DeCaf using the Random Forest model was 4x times the baselines. In terms of precision, as we can see in Table \ref{comparison}, DeCaf has 9.1\% higher precision than the baseline. Overall, we can see that DeCaf gives significantly better performance in diagnosing performance issues from service logs.

\textbf{Runtime performance}: We evaluated the runtime performance of the DeCaf system on two datasets from the Orion service. The Large dataset contained \textasciitilde50GB of daily logs with \textasciitilde45 million rows, 51 attributes and cumulative cardinality of 1.3 million. Whereas, the Small dataset was \textasciitilde100MB in size, containing \textasciitilde0.1 million rows. We did the runtime analysis using the data from the production deployment running in the Cosmos MapReduce cluster. Table \ref{runtime-comparison} shows the runtime of different stages of DeCaf. As we can see, data pre-processing and model training are the most expensive task in terms of runtime. The rest of the steps are near real-time with \textasciitilde1 min runtime. The diagnosing and triaging stages have similar runtime for both the datasets, because these stages do not operate on the raw dataset but the model output and the 14 day historical results. The triaging stage is comparatively slower than the diagnosing stage because of the multiple SQL queries which need to be run in the cloud in order to compare with historical results.

%% file: discussion.tex
\section{Discussion}
%related work at the end???
% 20 hours 
In this section, we first discuss the generalizability of DeCaf to other services and then discuss some of the future work.

\subsection{Generalizability}
All commercial large scale services have commitments to their customers for meeting SLOs for their KPIs. They also produce large amount of logs and telemetry which makes it prohibitively expensive to root cause and triage any regression in the KPIs. We designed and built DeCaf after looking at the requirements and existing DevOps practices of several services within Microsoft. As our case studies show, DeCaf has significantly reduced the manual effort for the on-call engineers. We believe, DeCaf solves a common problem which is applicable to almost any large scale service. It can be used to not just diagnose known incidents but also find existing performance issues before they lead to widespread customer impact. Further, to deploy DeCaf, service owners only need existing domain knowledge about the service logs and a one time manual effort of writing the data aggregation queries and hyper-parameter tuning.

\subsection{Future Work}
Today, even though DeCaf can handle heterogeneous attributes, it requires the data to be in a structured format. One of the next steps, will be to automate the transformation of the unstructured logs so that DeCaf can operate directly on the raw data. We can use unsupervised machine learning techniques to extract key-value pairs for various attributes from the logs. Further, DeCaf currently requires service owners to do feature selection based on domain knowledge. We believe, this step can be automated based on past incident and root cause knowledge bases. We can leverage NLP techniques to learn which attributes in the logs are useful. This information can then be surfaced to the service owners as recommendations if not to completely eliminate the manual effort. Lastly, another interesting direction will be to diagnose multiple KPIs simultaneously instead of diagnosing individual KPI separately.

%% file: conclusion.tex
\section{Conclusion}
We have described DeCaf, a system for diagnosing and triaging performance regressions in large-scale cloud services. Terabytes of logs and the heterogeneous attributes generated by large-scale cloud services makes it infeasible to do any automated or manual analysis. DeCaf leverages Random Forest models along with custom scoring functions to mine predicates from the logs which are correlated with regressions. Furthermore, the results are automatically triaged, making it easier for the on-call engineers to prioritize and investigate these issues. DeCaf has been deployed for two large-scale commercial cloud services in Microsoft. It was able to diagnose 10 known and 31 unknown performance issues while significantly reducing manual effort. We have also shared learnings and insights from integrating DeCaf into the DevOps processes at Microsoft.

%% file: acknowledgements.tex
\section{Acknowledgments}
We would like to acknowledge the invaluable contributions and support of B. Ashok, Pradnya Kulkarni, Kushal Narkhede, Nachiappan Nagappan, Jim Kleewein, Ranjita Bhagwan and Saravanakumar Rajmohan.